\documentclass[aps,prd,showpacs,nofootinbib,superscriptaddress,amsmath,amssymb]{revtex4}
\usepackage{amsfonts}
\usepackage{bm}
\usepackage{verbatim}

\usepackage{graphicx}
\usepackage{graphics}
\usepackage{color}

\newcommand{\revision}[1]{{#1}}
\newcommand{\revisionK}[1]{{#1}}

\usepackage{epsfig}

\begin{document}

\title{Anatomy of the chiral vortical effect}

\author{Ruslan Abramchuk}
\affiliation{Moscow Institute of Physics and Technology, 9, Institutskii per., Dolgoprudny, Moscow Region, 141700, Russia}

\author{Z.V.Khaidukov}
\affiliation{Institute for Theoretical and Experimental Physics, B. Cheremushkinskaya 25, Moscow, 117259, Russia}

\author{M.A. Zubkov \footnote{On leave of absence from Institute for Theoretical and Experimental Physics, B. Cheremushkinskaya 25, Moscow, 117259, Russia}}
\email{zubkov@itep.ru}
\affiliation{Physics Department, Ariel University, Ariel 40700, Israel}

\date{\today}

\begin{abstract}
We consider the system of relativistic rotating fermions in the presence of rotation. The rotation is set up as an enhancement of the angular momentum. In this approach the angular velocity for the angular momentum plays the same role as the chemical potential for density. We calculate the axial current using the direct solutions of the Dirac equation with the MIT bag boundary conditions. Next, we consider the alternative way of the rotation description, in which the local velocity of the substance multiplied by the chemical potential serves as the effective gauge field. In this approach this is possible to relate the axial current of the chiral vortical effect for the massless fermions to the topological invariant in momentum space, which is robust to the introduction of interactions. We compare the results for the axial current obtained using the two above mentioned approaches.
\end{abstract}

\maketitle

\section{Introduction}
\label{SectIntro}

Chiral vortical effect is the appearance of axial current in fermionic system in the presence of rotation. This effect was predicted for the first time by Vilenkin in \cite{Vilenkin}. From \cite{Vilenkin} the following expression for the axial current of massless Dirac particles may be read off (in the limit of high temperatures):
\begin{eqnarray}
\vec{j}^5=\frac{1}{6}\vec{\Omega}T^{2} \label{eq0}
\end{eqnarray}
%\section{Chiral vortical effect through the direct solution of Dirac equation in the presence of rotation}
In the presence of chemical potential (see for example  \cite{Volovik2003,chiralhydro,chernodub} and references therein) the additional term arises in the expression for the axial current:
\begin{eqnarray}
    \vec{j}^{5}=\Big(\frac{T^{2}}{6}+\frac{\mu^2}{2\pi^2}\Big)\vec{\Omega} \label{eq1}
\end{eqnarray}
There was a hope, that the value of the coefficient in front of vorticity does not depend on the interactions and can be fixed by the chiral anomaly, which is not subject to such corrections. It was shown, however \cite{Corr1,Corr2}, that the higher orders of perturbation theory are able to correct, at least, the coefficient in front of $T^2$ in Eq. (\ref{eq1}).

In the presence of the finite mass of the fermions the expression for the chiral vortical effect is changed as well as the expression for the chiral separation effect \cite{Metl}. On the formal level the theory with massless fermions suffers from various infrared divergencies \cite{infrared1,infrared2,infrared3}, for which reason the finite fermion masses are always introduced to the considered system even if the limit of small masses is assumed. Notice, that in \cite{Vilenkin} the neutral particles were discussed, and for them there are no infrared divergencies related to the radiation of photons.

It is worth mentioning, that the chiral vortical effect may be relevant for the description of the quark matter under extreme conditions in the rotated neutron stars~\cite{Cook:1993qr}, and the rotated fireballs that appear in the non - central heavy ion collisions~\cite{ref:HIC}. It has been argued that in the latter case the quark matter exists in the quark - gluon plasma phase \cite{QCDphases,1,2,3,4,5,6,7,8,9,10}.

Free rotating fermions ~\cite{Iyer:1982ah,ref:Becattini,Ambrus:2014uqa,Ambrus:2015lfr,Manning:2015sky} as well as the interacting systems in the presence of rotation \cite{chernodub} have been investigated extensively during the recent years. Rotation of the relativistic system may be considered only in the finite region of space ~\cite{Ambrus:2014uqa,ref:superluminal,Chernodub:2017ref}. The interplay of rotation and the magnetic field directed along the rotation axis has been considered, for example, in ~\cite{Chen:2015hfc}.

The chiral vortical effect belongs to the  family of the so - called anomalous transport phenomena. Such phenomena have also their incarnations in the solid state physics ~\cite{ref:Weyl}. This occurs, in particular, because the electronic system of the discovered recently Weyl and Dirac semimetals simulates relativistic physics and the corresponding excitations at the low energies are described by  Dirac equation \cite{semimetal_discovery,semimetal_discovery2,semimetal_discovery3,semimetal_effects6,semimetal_effects10,semimetal_effects11,semimetal_effects12,semimetal_effects13,Zyuzin:2012tv,tewary,16}. The other representatives of this family are, for example, the so - called anomalous quantum Hall effect \cite{AQHE}, the chiral separation effect \cite{Metl}, the chiral magnetic effect \cite{Kharzeev:2009pj}, the scale magnetic effect \cite{Chernodub:2016lbo}, and the rotational Hall effect \cite{Z2018}. Recently the presence of the anomalous quantum Hall effect was predicted in the three -- dimensional systems: the Weyl semimetals \cite{Zyuzin:2012tv,tewary} and the topological insulators \cite{Z2016_1}. At the same time the absence of the equilibrium version of the chiral magnetic effect was proved in \cite{Z2016_2}. 
It is expected, that both the chiral separation effect and the chiral vortical effect as well as the proposed recently rotational Hall effect \cite{Z2018} are expected to be observed in the heavy ion collisions \cite{ref:HIC}.

In the present paper we concentrate on the chiral vortical effect and discuss certain modifications of Eq. (\ref{eq1}). One of the sources of the corrections to Eq. (\ref{eq1}) is the finite size of the system. Rotation velocity cannot exceed the speed of light. Therefore, the following relation must be satisfied:
\begin{eqnarray}
\Omega R \le 1
\end{eqnarray}
where $\Omega$ is the angle velocity while $R$ is the radius of the system. 
In this way the additional parameter arises, which corresponds to the infrared physics. The relationship between temperature, chemical potential and the size of the system determines the behaviour of the system in the absence of interactions. The important effect, related to the finite size of the system, is the existence of the edge states, that can be localized on the boundary and that contribute the total axial current.

The possible modifications of Eq. (\ref{eq1}) due to the finite size corrections, interactions and the finite fermion mass constitute the "anatomy of the chiral vortical effect". Some of those issues will be discussed in the present paper. First, we will consider the non - interacting fermions and investigate the influence on the axial current of both the finite mass of the fermions and of the finite size of the system. The rotation of the system will be introduced as the enhancement of the total angular momentum. We adopt the MIT  boundary conditions in order to consider the system in the finite volume. At the same time the results of \cite{Vilenkin} were obtained in the infinite volume limit. Generally speaking, our results for the density of the axial current on the rotation axis reproduce those of \cite{Vilenkin} and \cite{chiralhydro}. Far from the rotation axis the effect is changed drastically. To the best of our knowledge this modification of the chiral vortical effect due to the finite volume has been considered here for the first time.

Besides, following \cite{chiralhydro} we will consider the alternative definition of rotation, where the four velocity of substance multiplied by the chemical potential is considered as the effective $U(1)$ gauge field.
This allows to reduce the chiral vortical effect to the chiral separation effect caused by the  corresponding effective magnetic field. The previous consideration of the latter effect \cite{KZ2017} in the infinite volume limit at zero temperature ensures that it is protected topologically, which means that the coefficient in the $\sim \mu^2$ term of Eq. (\ref{eq1}) should not be renormalized via interactions. 
 
The paper is organized as follows: in Section 2 we discuss  Dirac equation  in the presence of rotation that is introduced as the enhancement of the angular momentum, in  Section 3 we present the solutions of the Dirac equation. In  Section 4  we  discuss the constraints on those solutions that follow from the boundary conditions. In Section 5 we describe the calculation of axial current.
In Section 6 we adopt the alternative definition of rotation through the effective $U(1)$ gauge field and apply the previously known expressions for the chiral separation effect to the chiral vortical effect for the massive fermions.  In section 7 we discuss our numerical results and compare the two mentioned above approaches to the definition of rotation. In Section 8 we end with the conclusions.

\section{Rotation as the enhancement of angular momentum}
\label{SectOM}

Let us consider the description of rotation, which is due to the enhancement of the angular momentum in the theory by the term in the action proportional to the projection of momentum to the given axis:
$$
\delta S = \kappa \int d t \omega_{ij} M^{ij}
$$
where $\omega_{ij}$ marks the rotation plane (for the rotation around the $z$ axis $\omega_{ij} = \epsilon_{03ij}$) while $M^{ij}$ is the momentum tensor of the Dirac field:
$$
M^{ij} = \frac{1}{2}\int d^3 x \bar{\psi}\Big(\gamma^0\{ x^i ,\hat{P}^j\} -\gamma^0 \{x^j, \hat{P}^i\} + \{\gamma^0 , \frac{1}{2}\Sigma^{ij}\}\Big)\psi
$$
Thus the rotation around the $z$ axis on the level of the Dirac equation may be described by the following modification:
 \begin{equation}
        \Big[\gamma^0\mu  + \gamma^0\kappa \frac{\epsilon_{03ij}}{2}\Big(i\{ x^i ,\partial^j\} -i \{x^j, \partial^i\} +  \Sigma^{ij}\Big) + i \gamma^\mu \partial_\mu-M\Big]\psi=0 \label{direq2}
     \end{equation}
that is (using the Weyl basis)
\begin{equation}
    \begin{pmatrix} -M & i\partial_t + \mu + \kappa\left(i x^1 \partial^2 -i x^2 \partial^1 + \frac{1}{2}\sigma^{3}\right)- i\sigma^k\partial_k \\
    i\partial_t + \mu + \kappa\left(i x^1 \partial^2 -i x^2 \partial^1 + \frac{1}{2}\sigma^{3}\right) + i\sigma^k\partial_k & -M \end{pmatrix}
    \psi = 0 \label{direq3}
\end{equation}
Formally this equation is equivalent to Eq. (2.17) of \cite{chernodub} if we identify $\kappa$ with angular velocity $\Omega$. The description of rotation in the rotating reference frame is equivalent to the one in the laboratory reference frame if the latter is understood as an enhancement of the angular moment.

We rewrite the Dirac equation using the expression for the total angular momentum \(\hat{J}_z=\hat{L}_z+\frac{1}{2}\Sigma^{12}\)  and the notation \(\hat{P}_\pm = \hat{p}_x\pm i\hat{p}_y\)
\begin{equation}
    \begin{pmatrix} -M & W + \mu + \kappa \hat{J}_z- \sigma^3k -
        \begin{pmatrix} & \hat{P}_- \\ \hat{P}_+ & \end{pmatrix} \\
    W + \mu + \kappa\hat{J}_z + \sigma^3k +
        \begin{pmatrix} & \hat{P}_- \\ \hat{P}_+ & \end{pmatrix} & -M \end{pmatrix} \psi = 0 \label{direq4}
\end{equation}
Notice, that in the cylindrical coordinates
\begin{equation}
    \hat{P}_\pm = -ie^{\pm i\phi}(\partial_r \pm ir^{-1}\partial_\phi)
\end{equation}

\section{Solutions of the Dirac equation}

The helicity operator and the total angular momentum commute
\begin{equation}
    \left[\hat{h},\hat{L}_z+\frac{1}{2}\sigma_z\right] = 0,\quad \hat{h} = \frac{\bf{\sigma}\bf{p}}{p}\label{comm}
\end{equation}
Therefore, let us look for the solutions with definite helicity \(\sigma=\pm 1\) and total angular momentum \(j_z=m+\frac{1}{2}\)
\begin{equation}
    \psi_j = e^{-i\revision{W}t+kz}\begin{pmatrix} C^L_j\varphi_j(\rho,\phi) \\ C^R_j\varphi_j(\rho,\phi) \end{pmatrix}
        ,\quad j = (\revision{W},k,m,\sigma) \label{helicalst}
\end{equation}
where \(\varphi_j\) is the (two-component)  eigenfunction of the helicity operator
\begin{equation}
    \hat{h}\varphi_j = \sigma\varphi_j
\end{equation}
We will use the following relations for \(\hat{P}_\pm\)
\begin{equation}
    P_\pm e^{im\phi}J_m(qr) = \pm iqe^{i(m\pm 1)\phi}J_{m\pm 1}(qr)
\end{equation}
We assume \(q\ge 0\), the opposite sign would correspond to the same eigenfunction. Here \(J_m\) is the Bessel function of order \(m\). We are
 looking for \(\varphi\) of the following form
\begin{equation}
    \varphi = \begin{pmatrix}A e^{im\phi}J_m(qr) \\ B e^{i(m+1)\phi}J_{m+1}(qr)\end{pmatrix}
\end{equation}
and obtain
\begin{equation}
    \begin{pmatrix}
        k-\sigma p & -iq \\ iq & -k-\sigma p
    \end{pmatrix}
    \begin{pmatrix} A\\B \end{pmatrix} = 0
\end{equation}
This gives
\begin{equation}
    p =\sqrt{ k^2+q^2}
\end{equation}
and
\begin{equation}
    \begin{pmatrix}A\\B\end{pmatrix} = \frac{1}{\sqrt{2}p}
    \begin{pmatrix}\revision{iq}\\k-\sigma p\end{pmatrix} \label{coef}
\end{equation}
As a result \eqref{direq4} is reduced to
\revision{
\begin{equation}
    \begin{pmatrix}-M & w-p_j\sigma \\
    w+p_j\sigma & -M \end{pmatrix}
    \begin{pmatrix}C^L\\C^R\end{pmatrix} = 0 \label{EQ}
\end{equation}
where
\begin{equation}
    w = W+\mu+\Omega j_z
\end{equation}
The following dispersion relation follows
\begin{equation}
    w = \pm\sqrt{k^2 + q^2 + M^2}
\end{equation}
The solution of Eq. (\ref{EQ}) is given by
\begin{equation}
    \begin{pmatrix}C_L\\C_R\end{pmatrix} =
        \frac{1}{M}\begin{pmatrix}w-p\sigma \\ M \end{pmatrix}
\end{equation}
}
As expected, at the vanishing mass the helicity and the chirality cannot be separated:
\(C_L=1,\,C_R=0,\,\sigma=-1\) or \(C_L=0,\,C_R=1,\,\sigma=1\).

\section{The MIT bag conditions}

In order to avoid the infrared singularities we place the system to the cylinder of radius $R$ and imply the so - called MIT bag boundary conditions
\begin{equation}
    (i\gamma^\mu n_\mu -1)\psi\rvert_{r=R} =
    \begin{pmatrix}-1&-i\begin{pmatrix}&e^{-i\phi}\\e^{i\phi}&\end{pmatrix}\\
        i\begin{pmatrix}&e^{-i\phi}\\e^{i\phi}&\end{pmatrix}&-1
    \end{pmatrix}
    \psi\rvert_{r=R} = 0 \label{mitbc}
\end{equation}
here $n_\mu = (0, \frac{\bf r}{r}, 0)$ is the unit vector orthogonal to the cylinder surface.
\revision{According to those conditions the current normal to the surface of the cylinder vanishes:
\[
j^\mu n_\mu = 0
\]}
This boundary condition mixes the states \revisionK{\(\psi_\sigma\), given by Eq. \eqref{helicalst},} with the opposite helicities \revisionK{\(\sigma=\pm1\)}
\begin{equation}
    \psi = C_+\psi_+ + C_-\psi_- \label{ans}
\end{equation}
The non-zero \(C_\pm\) exists if \(q = q_{ml}\),
where the radial quantum number $l$ enumerates the admitted values of $q$ at the given values of \(m\). It
satisfies the following quantization condition
\begin{equation}
    {\bf j}_m^2-\frac{2M}{q}{\bf j}_m-1=0,\quad {\bf j}_m(q) = \frac{J_m(qR)}{J_{m+1}(qR)} \label{Jq}
\end{equation}
This gives
$$
    {\bf j}_m(q) = \frac{M}{q}\pm\sqrt{\left(\frac{M}{q}\right)^2+1}
$$
Notice, that at $q=0$ the wavefunction becomes trivial $\psi\equiv 0$
while at $ MR\ll 1$ we have $q_{m1}R \approx \alpha_{m1}$, where $J_m(\alpha_{ml})=J_{m+1}(\alpha_{ml})$.

Notice, that at the certain value of mass the lowest energy state disappears, i.e. if $MR=m+1\text{ for }m\ge 0\text{ or }MR=-m\text{ for }m<0$ then $ q_{m1}=0$. However, at the large values of masses $MR\gg 1$ we get
$q_{m1}R\approx \gamma_{m1}-0$ where $J_m(\gamma_{ml})=0$ at $m>0$ and $J_{m+1}(\gamma_{(m+1),l})=0$ at $m\le 0$.

It is worth mentioning, that Eq. (\ref{Jq}) has the imaginary roots \(q=i\nu,\,\nu\ge 0\) \cite{edgestates}.
Let us rewrite it using the modified Bessel functions \(I_m(x)=i^{-m}J_m(ix)\):
\begin{equation}
    {\bf i}_m^2-\frac{2M}{\nu}{\bf i}_m+1=0,\quad {\bf i}_m(\nu) = \frac{I_m(\nu R)}{I_{m+1}(\nu R)}
\end{equation}
Its solution gives
$$
    {\bf i}_m(\nu) = \frac{M}{\nu}+\sqrt{\left(\frac{M}{\nu}\right)^2-1}
$$
At $MR \gg 1$ we obtain $\nu \approx M-0$, i.e. $M^2-\nu_m^2\approx\left(\frac{j_z}{R}\right)^2$.
At $MR \ll 1$ we come to $\nu_m = 0$
while $\nu_m > 0 $  if  $MR>m+1$ for $m>0$ or $MR>-m$ for $m<0$.
\(I_m(x)\ge 0\) and grows exponentially, so that there is only one solution \(\nu_m\) and one can say, that the corresponding solution of the Dirac equation represents the edge state
(as opposed to the bulk states). It is localized at the edge of the cylinder.
 In the massless limit \(\nu_m=0\) for \(m\ne 0\) while \(\nu_0\) does not exist, which means that there are no edge states.

Now the solution of \revision{Eq. (\ref{mitbc}) is given by
\begin{equation}
    C_\pm = (\pm w+p)qJ_m(qR) + M(\pm k+p)J_{m+1}(qR)
\end{equation}
}

\revision{
    In the massless limit these boundary conditions assume the existence of both the \revisionK{right}  - handed particles and the \revisionK{left} - handed  anti - particles.
 One may define the boundary conditions in a more general way introducing the ``chiral angle'' \(\theta\)
\begin{equation}
    (i\gamma^\mu n_\mu -e^{i\theta})\psi\rvert_{r=R} = 0
\end{equation}
which alters the spectrum. In particular, at \(\theta=\pi\) the particles are to be \revisionK{left} - handed while the antiparticles are \revisionK{right} -  handed. However, below we restrict ourselves to the case \(\theta=0\).
}

We normalize those solutions as follows (``one particle in the volume'')
\begin{equation}
    \int d^3x \psi_{kml}^\dag \psi_{k'm'l'} =  \frac{2\pi\delta(k-k')}{L_z}\delta_{mm'}\delta_{ll'} \label{norm}
\end{equation}

\begin{figure}
    \includegraphics[width=0.7\linewidth]{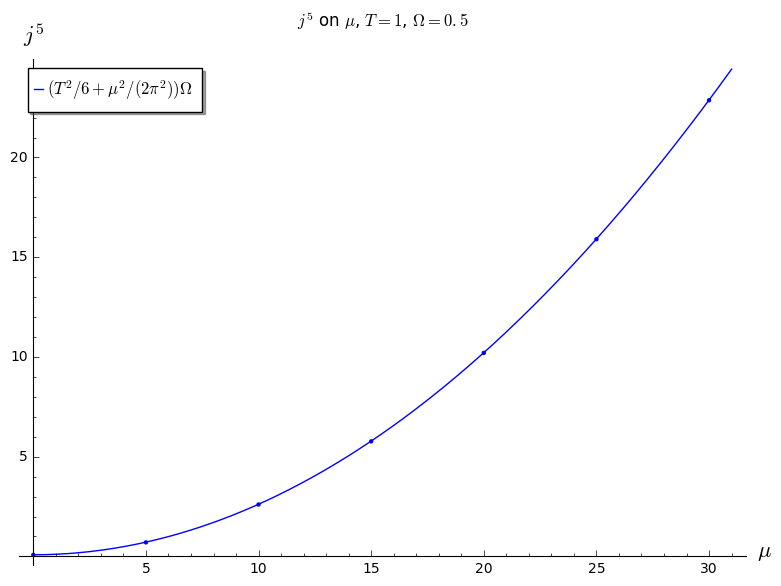}
    \caption{Axial current density at $r=0$ for the system of massless fermions rotating with the angular velocity $\Omega = 0.5 /R$. Temperature is taken equal to $T = 1/R$. The values of chemical potential $\mu$ are presented in the units of $1/R$.}
    \label{f1}
\end{figure}

\begin{figure}
    \includegraphics[width=0.7\linewidth]{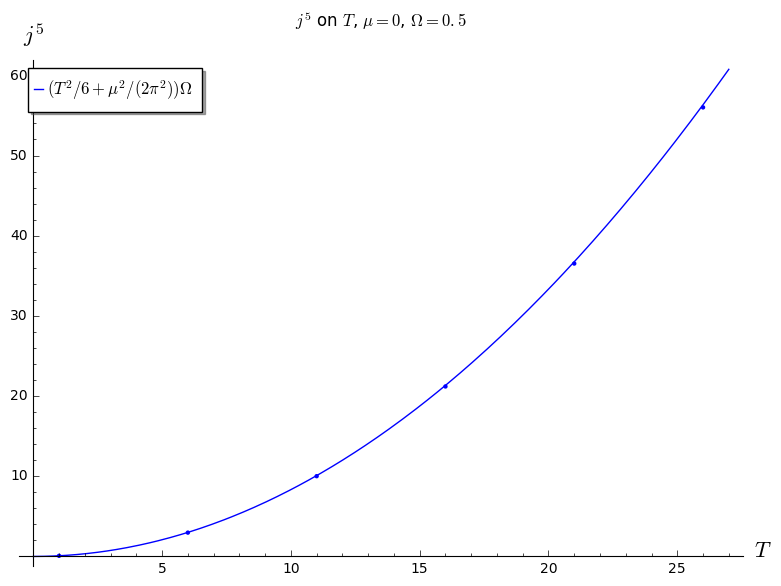}
        \caption{Axial current density at $r=0$ for the system of massless fermions rotating with the angular velocity $\Omega = 0.5 /R$. The chemical potential is equal to zero. The values of temperature $T$ are presented in the units of $1/R$.}
        \label{f2}
\end{figure}

\begin{figure}
    \includegraphics[width=0.7\linewidth]{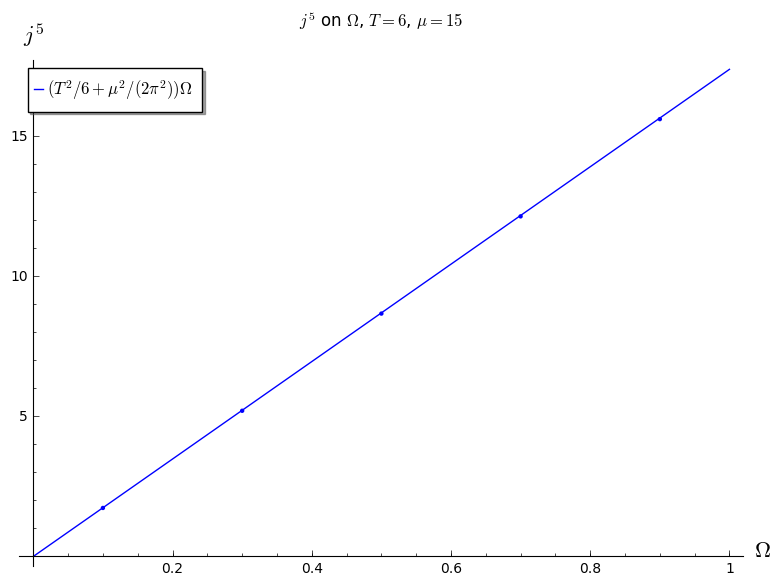}
    \caption{Axial current density at $r=0$ for the system of massless fermions rotating with the angular velocity $\Omega$ (the values of $\Omega$ are represented in the units of $1/R$). Temperature is taken equal to $T = 6/R$. The value of chemical potential is $\mu = 15/R$.}
    \label{f3}
\end{figure}

\section{Calculation of the axial current}

We are interested in the axial current along the rotation axis
\begin{equation}
    j^5_z = \bar{\psi}\gamma^5\gamma^3\psi = \psi^\dag_L\sigma^3\psi_L+\psi^\dag_R\sigma^3\psi_R
\end{equation}
The Fermi distribution is given by
\begin{equation}
    n(w,j_z) = \frac{1}{e^{\beta(w - \mu - \Omega j_z)}+1}
\end{equation}
This distribution describes not only the positive energy states ($E=w-\revision{\Omega}j_z > 0$), but also the states with the negative energy $E = w - \revision{\Omega}j_z <0$.

In the following we will use the assumption that in the presence of rotation the vacuum (i.e. the occupied states with negative energy) gives vanishing contribution to the total axial current. Therefore, for $E = w - j_z \revision{\Omega} <0$ we change the Fermi distribution to \begin{equation}
    n(w,j_z)_{w-j_z \revision{\Omega} <0} = \frac{1}{e^{\beta(w - \mu - \Omega j_z)}+1} - 1 = -\frac{1}{e^{-\beta(w - \mu - \Omega j_z)}+1}
\end{equation}
Overall, we come to the following version of the Fermi distribution to be used for the calculation of the axial current:
\begin{equation}
    n(w,j_z) = \frac{{\rm sign}\,(w-j_z \Omega )}{e^{\beta(w - \mu - \Omega j_z){\rm sign}\,(w-j_z \Omega )}+1}\label{nOmega}
\end{equation}

We may change the above procedure for the practical calculations because for all considered above eigenstates of energy the following relation is valid
$$
{\rm sign}\,(w-j_z \Omega ) = {\rm sign}\,(w)
$$
It has been checked numerically for the considered values of parameters.
 Then  one may take
\begin{equation}
    n(w,j_z) = \frac{{\rm sign}\,(w )}{e^{\beta(w - \mu - \Omega j_z){\rm sign}\,(w)}+1}
\end{equation}
We have the final expression for the chiral current density
\begin{equation}
    \langle j^5_z(r)\rangle_\beta = \sum_{k,q,{\rm sign}\,(w),j_z} n(w,j_z) \bar{\psi}_{k,q,{\rm sign}\,(w),j_z}\gamma_5\gamma_3\psi_{k,q,{\rm sign}\,(w),j_z} \label{j5avg}
\end{equation}
Here the sum is over all quantum numbers, which enumerate the eigenstates of energy. Here by $\psi_{k,q,{\rm sign}\,(w),j_z}$ we denote the corresponding eigenfunction, which is the $4$ - component complex - valued spinor. Correspondingly, $\bar{\psi} = (\psi^*)^T \gamma^0$ is the conjugated $4$ - component spinor.

In \cite{Vilenkin} the analytical expression for the axial current has been obtained for the case of vanishing mass. Here in addition we obtain the analytical expression for $j^5_z(0)$ at nonzero $M$ but for $\mu = 0$:
\begin{equation}
    \langle j^5_z(r)\rangle_\beta = \frac{2\Omega T^2}{\pi^2} \int_{M/T}^{\infty} \frac{zdz}{e^z+1} \label{JT}
\end{equation}

\revisionK{
We expect that at low temperature \(T<R^{-1}\) and zero chemical potential  there are no excited states and, therefore, the axial current density vanishes. The dependence of axial current density on the chemical potential at low temperatures reflects directly the details of spectrum.
}

\revisionK{
Let us analyze the axial current density on the rotation axis in the massless limit.
The current density is saturated by the states with the minimal total angular momentum \(j_z=\pm 1/2\).
At the temperatures \(T<\Omega\) the states with the opposite projections of the total angular momentum are excited when the chemical potential grows, and this does not occur simultaneously.
}

\revisionK{
For simplicity, let us discuss the case \(T=0\).
Then Fermi distribution becomes the step-function, and for \(0<\mu<q_{02}-\Omega/2\)  only the two states with the minimal transverse momentum \(q\equiv q_{01}\) and \(j_z=\pm 1/2\) may be excited, so we are able to directly use Eq. \eqref{j5avg}.
Corresponding states (up to an irrelevant here phase factor)
\begin{equation}
    \psi_{j_z=1/2}=N_{1/2}pqJ_0(qR)\begin{pmatrix}0\\0\\iqJ_0(qr)\\(k-p)e^{i\phi}J_1(qr)\end{pmatrix}
    ,\quad
    \psi_{j_z=-1/2}=N_{-1/2}pqJ_0(qR)\begin{pmatrix}0\\0\\iqe^{-i\phi}J_{-1}(qr)\\(k-p)J_0(qr)\end{pmatrix}
\end{equation}
We normalize the states according to Eq. \eqref{norm} using the following relation and the transverse momentum quantization conditions
\[
    \int_0^1zdzJ_n^2(\lambda z) = \frac{1}{2}(J_{n+1}^2(\lambda)-J_n^2(\lambda)) + \frac{n}{\lambda^2}J_n(\lambda)J_{n+1}(\lambda)
\]
and derive
\begin{equation}
    N_{1/2}^{-1} = \sqrt{L_z}J_0^2(qR)p(p-k)
    ,\quad
    N_{-1/2}^{-1} = \sqrt{L_z}J_0^2(qR)pq
\end{equation}
Using Eq. \eqref{j5avg}, up to a normalization factor we obtain
\begin{equation}
    \langle j^5_z(0)\rangle \sim \int\frac{dk}{2\pi}\left[
    \frac{q^4}{(k-p)^2}\theta\left(\sqrt{\left(\mu+\frac{\Omega}{2}\right)^2-q^2}-|k|\right)
    -(k-p)^2\theta\left(\sqrt{\left(\mu-\frac{\Omega}{2}\right)^2-q^2}-|k|\right)\right]
\end{equation}
The function is zero up to \(\mu=q-\frac{\Omega}{2}\), then it grows, and has a peak at \(\mu=q+\frac{\Omega}{2}\), where the \(j_z=-1/2\) state get excited. After that it approaches a constant.
}

\revisionK{
The current density has a series of similar peaks at \(\mu \approx q_{0l}+\frac{\Omega}{2},\,l\in \mathbb{N}\).
The non-zero temperature smooths out the peaks and the distortion has the form of the oscillations.
}

\revisionK{
    We calculate the axial current density using Eq. \eqref{j5avg} directly.
    First, we find the transverse momenta \(q=q_{ml}\) (and \(q=i\nu_m\) if there are the edge states) by numerical evaluation of the roots of Eq. \eqref{Jq}. Then, we sum the series of Eq. \eqref{j5avg} up to the values of \(m,l\) large enough to provide the relative precision of the order of \(10^{-3}\). All integrations encountered during the computation of the terms in the series were performed numerically.
    To achieve the necessary precision we analytically estimate the sum of the series in the large \(q\) limit. Our procedure allows to obtain the error bars smaller than the markers of points on the presented figures.
}

\begin{figure}
    \includegraphics[width=0.7\linewidth]{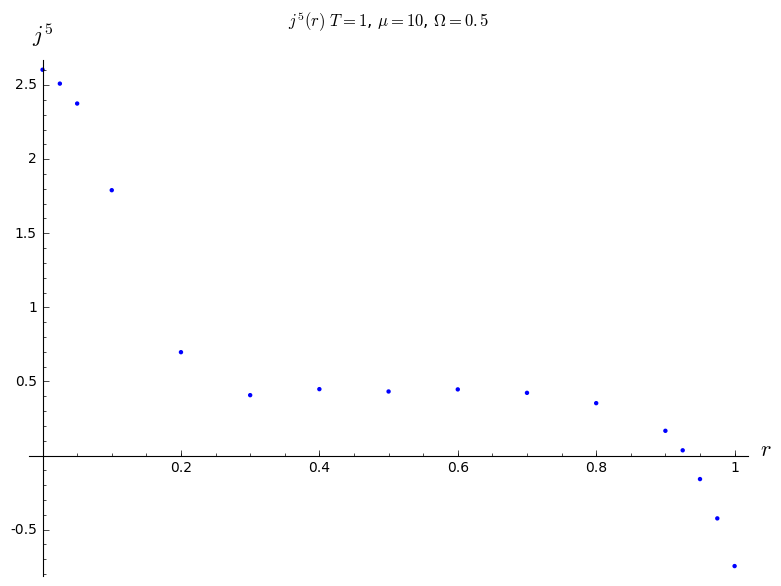}
    \caption{Axial current density as a function of $r$ (given in the units of $R$) for the system of massless fermions rotating with the angular velocity $\Omega = 0.5 /R$. Temperature is taken equal to $T = 1/R$. The value of chemical potential is $\mu = 10/R$. }
    \label{f4}
\end{figure}

\begin{figure}
    \includegraphics[width=0.7\linewidth]{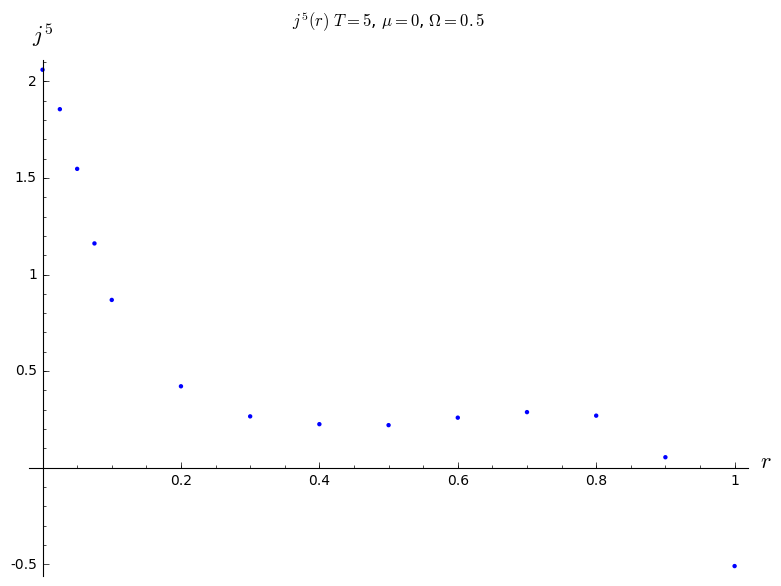}
        \caption{Axial current density as a function of $r$ (given in the units of $R$) for the system of massless fermions rotating with the angular velocity $\Omega = 0.5 /R$. Temperature is taken equal to $T = 5/R$. The value of chemical potential is $\mu = 0$.}
        \label{f5}
\end{figure}

\begin{figure}
    \includegraphics[width=0.7\linewidth]{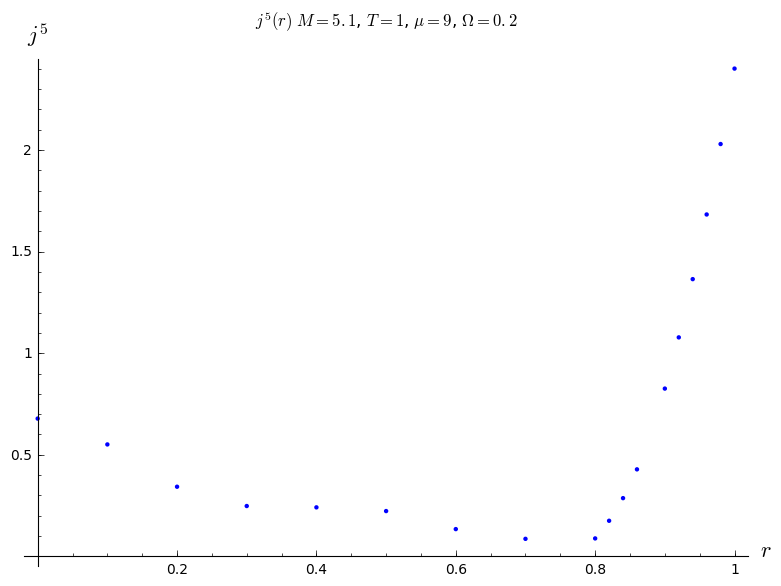}
        \caption{\revision{The axial current density (given in the units of $R^{-3}$) as a function of $r$ (given in the units of $R$) for the system of fermions with \(M=5.1R^{-1}\) rotating with the angular velocity $\Omega = 0.2R^{-1}$. Temperature is taken equal to  $T = 1R^{-1}$. The value of the chemical potential is $\mu = 9R^{-1}$.}}
        \label{f5A}
\end{figure}

\section{Description of rotation via the effective gauge field}
\label{SecFree}

\begin{figure}
    \includegraphics[width=0.7\linewidth]{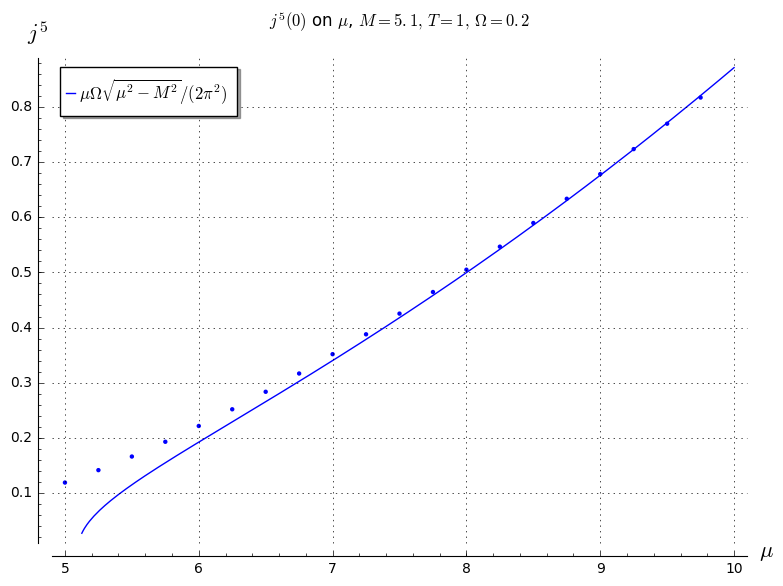}
        \caption{Axial current density at $r=0$ as a function of $\mu$ (given in the units of $1/R$) for the system of fermions with mass $M=5.1/R$ rotating with the angular velocity $\Omega = 0.2/R$. Temperature is taken equal to $T = 1/R$. }
        \label{f6}
\end{figure}

\begin{figure}
    \includegraphics[width=0.7\linewidth]{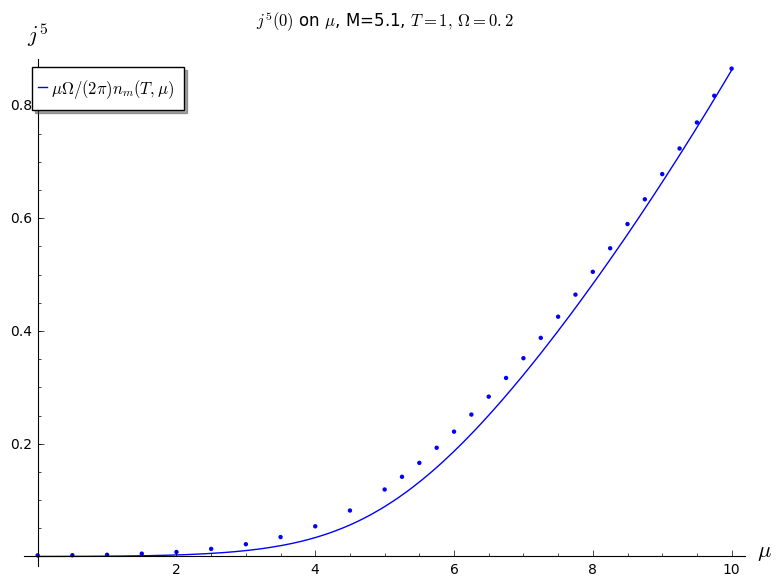}
        \caption{Axial current density at $r=0$ as a function of $\mu$ (given in the units of $1/R$) for the system of fermions with mass $M=5.1/R$ rotating with the angular velocity $\Omega = 0.2/R$. Temperature is taken equal to $T = 1/R$. }
        \label{f8}
\end{figure}

\begin{figure}
    \includegraphics[width=0.7\linewidth]{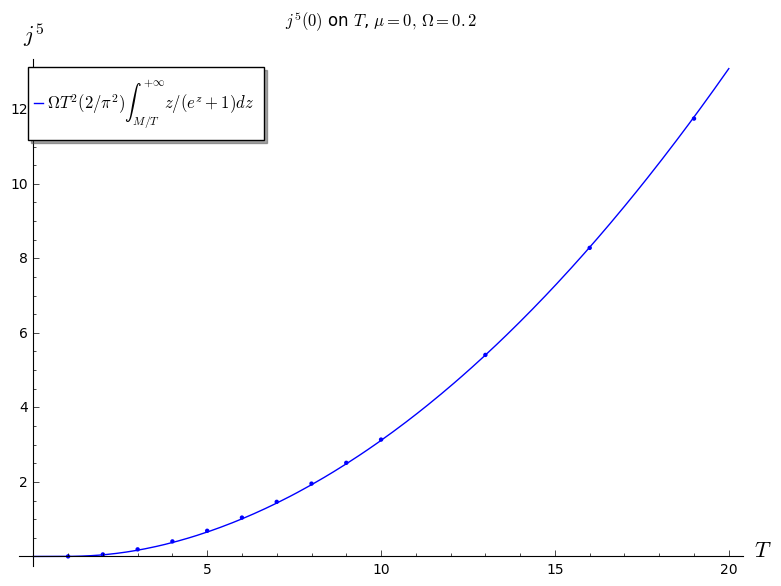}
        \caption{Axial current density at $r=0$ as a function of $T$ (given in the units of $1/R$) for the system of fermions with mass $M=5.1/R$ rotating with the angular velocity $\Omega = 0.2/R$. The value of chemical potential is taken equal to $\mu = 0$. }
        \label{f7}
\end{figure}

\begin{figure}
    \includegraphics[width=0.7\linewidth]{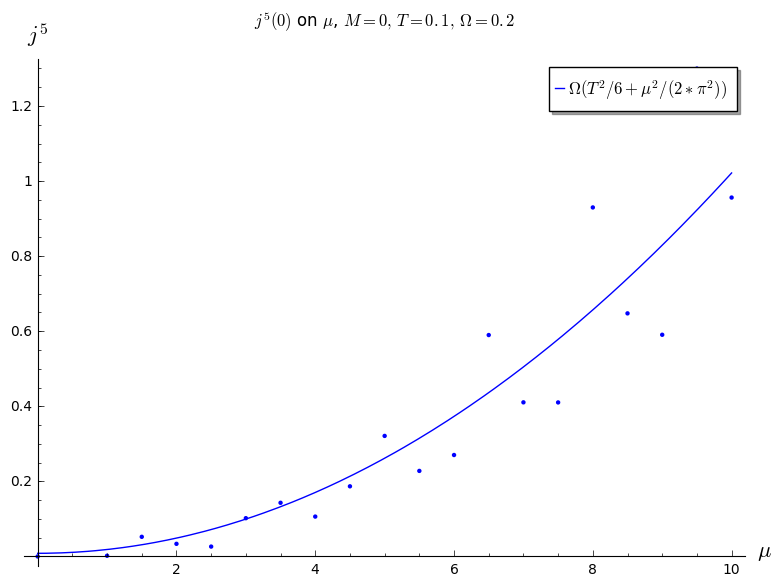}
        \caption{\revision{The axial current density at $r=0$ (given in the units of $R^{-3}$) as a function of $\mu$ (given in the units of $R^{-1}$) for the system of massless fermions rotating with the angular velocity $\Omega = 0.2R^{-1}$.}}
        \label{fig:osc}
\end{figure}

In this section we consider the alternative description of rotation of the fermionic matter with angular velocity \(\Omega\)  in the state with the chemical potential \(\mu\). Let us first neglect the interaction between the fermions. We require that the value of  $\Omega$ is so small, that the velocity $\Omega r$ never becomes larger than $1$ within the considered region of space (of radius $R$). Thus $\Omega$ and $R$  obey $\Omega_0 R <1$.
This is possible to describe the rotating substance in the small vicinity of a given point \cite{chiralhydro} by the action
\begin{equation}
        S = \int d^4x \bar{\psi}(\gamma^\mu(i\partial_\mu+\mu u_\mu)-M)\psi\label{action}
\end{equation}
where \(u_\mu\) is the four-velocity of the medium in the given point, which is defined as a unit four-vector tangent to a world-line of a piece of medium.  It can be obtained from the ordinary velocity of the piece of medium \(\vec{v}=(-\Omega y,\Omega x,0)^T\) as follows
        \begin{equation}
        u^\mu = \gamma(r)(1,-\Omega y,\Omega x,0)^T,\quad \gamma(r)=\frac{1}{\sqrt{1-\Omega^2 r^2}}
    \end{equation}
The corresponding Dirac equation is equivalent to the Dirac equation in the presence of the external Abelian gauge field potential $A_\mu = - \mu u_\mu$:
    \begin{gather}
        (\gamma^\mu(i\partial_\mu+\mu u_\mu)-M)\psi=0 \label{direq}
     \end{gather}
The space components of $u_\mu$ are equivalent to the effective magnetic field, which is directed along the axis of rotation and depends on $r$:
\begin{eqnarray}
    \mathbf{B} &=& -(\nabla\times(\mu \mathbf{u}(r))) \nonumber\\&& = -\mu \Omega \gamma(r) \Big(2+  \frac{d\,{\rm log}\,[ \mu \Omega \gamma(r)]}{d\,{\rm log}\,r}\Big)\mathbf{e}_z \label{Magn}
\end{eqnarray}
The $0$ - component of $u_\mu$ gives rise to the electric field with radial direction
\begin{gather}
    \mathbf{E} = \nabla \mu u_0(r) = \frac{d (\mu \gamma(r))}{dr}\frac{\mathbf{r}}{r}
\end{gather}
The particularly simple case is when the rotation velocity remains much smaller than unity, so that we may neglect the dependence of $\gamma(r)$ on $r$ and set $\gamma \approx 1$. Then the chemical potential in the laboratory reference frame $$\mu_{lab} = \mu \gamma(r)  \approx \mu$$ and we obtain $\mathbf{E} = 0$ while \begin{equation}
\mathbf{B} = (0,0, -2\mu \Omega ) \label{Beff}\end{equation}
In the following we will restrict ourselves by this case. Then we are left with the Dirac fermion in the presence of constant effective magnetic field $-2\mu \Omega$. The corresponding spectrum is discrete and consists of the Landau levels.

According to the results of \cite{KZ2017} the Chiral Separation effect gives the following expression for the axial current (assuming the fermions are massless):
\begin{equation}
j^{5k}= \frac{{\cal N} \epsilon^{ijk}}{4\pi^2} F_{ij} \mu\label{jmuH}
\end{equation}
with
\begin{eqnarray}
{\cal N}&=& \frac{1}{12}\int_{\Sigma}\frac{1}{(2\pi)^2}Tr \gamma^5 {\cal G}(\omega_{},\textbf{k})d{\cal
G}^{-1}(\omega_{},\textbf{k})\nonumber \\&&\wedge d{\cal G}(\omega_{},\textbf{k})\wedge {\cal
G}^{-1}(\omega_{},\textbf{k})\label{N_3}
\end{eqnarray}
where $\Sigma$ is the 3D hypersurface of infinitely small volume in momentum space that embraces the singularities of the Green function ${\cal G}(\omega_{},\textbf{k})$ concentrated at the Fermi points. This representation for the axial current is valid  and Eq. (\ref{N_3}) is  the topological invariant if $\gamma^5$ anti - commutes with the Green function in a small vicinity of the Fermi point. The advantage of this representation is that it is valid for the interacting system, in which we should use the complete fermion propagator $\cal G$. However, if this system may be transformed smoothly to the simple noninteracting one, then the value of ${\cal N}$ is not changed and may be calculated using the Green function of the latter system. In particular, for the system of one massless noninteracting fermion ${\cal N} = 1$.

In the presence of nonzero mass $M$ the situation is changed, and the poles of the Green function do not appear while $\mu < M$, which gives the vanishing CSE current. At $\mu \ge M$ the Fermi surface appears, and it contributes to the chiral current in a more complicated way. However, for the non - interacting fermion we have an expression derived in \cite{Metl}. For $T=0$ it has the form:
\begin{equation}
j^{5k}= \frac{\epsilon^{ijk}}{4\pi^2} F_{ij} \sqrt{\mu^2-M^2} \label{jmuH2}
\end{equation}
For the nonzero temperature the current may be read off from Eqs. (36) , (37) of \cite{Metl}:
\begin{equation}
j^{5k}= \frac{\epsilon^{ijk}}{4\pi} F_{ij} n_m(\beta,\mu), \quad n_m(\beta,\mu) = \int \frac{d k}{2\pi} \Big(\frac{1}{e^{\beta(\sqrt{M^2 + k^2} - \mu )}+1} - \frac{1}{e^{-\beta(\sqrt{M^2 + k^2} - \mu )}+1} \Big) \label{T20}
\end{equation}
where $\beta = 1/T$.
In particular, for $\mu = 0$ we arrive at
\begin{equation}
j^{5k}= \frac{\epsilon^{ijk}}{4\pi} F_{ij} \int \frac{dk}{2\pi} {\rm th}\Big(\frac{\beta \sqrt{M^2 + k^2}}{2}\Big) \label{T21}
\end{equation}

Following \cite{chiralhydro} we represent the axial current of the Chiral Vortical Effect as the axial current of the Chiral Separation Effect corresponding to the magnetic field of Eq. (\ref{Beff}). This gives for the system of the interacting massless fermions rotating with angular velocity $\Omega$ around the z axis:
 \begin{equation}
j^{5k}= \frac{{\cal N} \epsilon^{12k}}{2\pi^2}  \mu^2 \Omega \label{jmuH_}
\end{equation}
where ${\cal N}$ is the topological invariant given by Eq. (\ref{N_3}).
At the same time for the system consisting of one noninteracting massive fermion at zero temperature we have
\begin{equation}
    j^{5k}= \frac{ \epsilon^{12k}}{2\pi^2} \revisionK{|\mu|} \sqrt{\mu^2-M^2} \Omega\label{jmuH2_}
\end{equation}
For the nonzero temperature we obtain
\begin{equation}
j^{5k}= \frac{ \epsilon^{12k}}{2\pi^2} \mu \Omega n_m(\beta,\mu)\label{Jn}
\end{equation}
with
$$ n_m(\revision{T},\mu) = \int \frac{dk}{2\pi} \Big(\frac{1}{e^{\beta(\sqrt{M^2 + k^2} - \mu )}+1} - \frac{1}{e^{-\beta(\sqrt{M^2 + k^2} - \mu )}+1} \Big) $$
For $\mu = 0$ and nonzero $T$ Eq. (\ref{T21}) would give us the vanishing value of axial current contrary to Eq. (\ref{JT}). This means that Eq. (\ref{Jn}) does not approximate sufficiently well the result obtained within the model of Sect. \ref{SectOM} at small \revision{$\mu$, unless \(\mu-M\gg T\)}.

\section{Numerical results}

In this section we discuss our numerical results obtained for the axial current in the model of Sect. \ref{SectOM}. Notice, that in the presented plots the error bars are smaller than the sizes of the symbols that represent the  data.

\begin{enumerate}

\item{}

First of all let us consider the case of vanishing mass $M=0$. We expect that close to the rotation axis the influence of boundary may be neglected and therefore the results of \cite{Vilenkin} are to be reproduced. We will see that this indeed occurs (at least, for the values of temperature of the order of $1/R $ or larger). We illustrate our results by Fig. \ref{f1}, \ref{f2}, \ref{f3}.

On Fig. \ref{f1} the value of the axial current at $r=0$ is represented as a function of $\mu$. The system of massless fermions is rotating with the angular velocity $\Omega = 0.5 /R$. Temperature is taken equal to $T = 1/R$. The values of chemical potential $\mu$ are presented in the units of $1/R$.

On Fig. \ref{f2}  the axial current density at $r=0$ for the system of massless fermions rotating with the angular velocity $\Omega = 0.5 /R$ is represented as a function of $T$. The chemical potential is equal to zero. The values of temperature $T$ are presented in the units of $1/R$.

On Fig. \ref{f3} the axial current density at $r=0$ for the system of massless fermions is represented as a function of the angular velocity $\Omega$ (the values of $\Omega$ are given in the units of $1/R$). Temperature is taken equal to $T = 6/R$. The value of chemical potential is $\mu = 15/R$.

\item{}

The finite volume effects become relevant at the values of $r$ with finite ratios $r/R$. Our results are illustrated by Fig. \ref{f4}, Fig. \ref{f5} \revision{and Fig. \ref{f5A}}, where the profile of $j_5(r)$ is represented as a function of $r$ (given in the units of $R$). One can see, that the finite volume effects are relevant not only close to the boundary itself $R=1$ but at any finite nonzero value of the ratio $r/R$.

\revision{
The axial current density outside of the rotation axis differs essentially from its value on the axis itself. Everywhere the axial current density is saturated by the states with the transversal momentum \(q_{ml}\sim\sqrt{T^2+\mu^2}\).
However, on the rotation axis the states with the angular momentum \(j_z=\pm 1/2\) only contribute to the current density.
 Far out from the rotation axis the current density is saturated by the states with the higher angular momenta.
}

\revision{
On Fig. \ref{f4} and Fig. \ref{f5} the axial current density is represented as a function of \(r\) in the massless limit for various values of angular velocity, temperature and chemical potential.
}

\revision{
On Fig. \ref{f5A} the axial current density is represented as a function of \(r\) in the case of finite mass.
For the massive fermions \(MR\gtrsim 1\) the maximum of the axial current density is shifted out of the rotation axis.
}

\item{} We expect that at nonzero fermion mass and small enough values of $\Omega$ and $T$ the axial current density at $r =0$ given by the model of Sect. \ref{SectOM} is close to the value predicted within the different model of Sect. \ref{SecFree} in Eq. (\ref{jmuH2_}) and Eq. (\ref{Jn}).
    This expectation is indeed justified by our numerical results. We illustrate this by Fig. \ref{f6} and \ref{f8}, where we represent the value of the axial current density at $r=0$ as a function of $\mu$ (given in the units of $1/R$). One can see, that at $\mu - M \gg T$ indeed the dependence of $j_5(0)$ on $\mu$ is given by Eq. (\ref{jmuH2_}) or Eq. (\ref{Jn}).

    At the same time at $\mu = 0$ we obtain the result presented in Fig. \ref{f7}, which differs from the vanishing prediction of Eq. (\ref{Jn}).  Here we represent the axial current density at $r=0$ as a function of $T$ (given in the units of $1/R$) for the system of fermions with mass $M=5.1/R$ rotating with the angular velocity $\Omega = 0.2/R$. The value of chemical potential is taken equal to $\mu = 0$. One can see, that these data are well fitted by Eq. (\ref{JT}).

\item
    \revision{At small temperature \(T<R^{-1}\) the finite-size effects reveal themselves via the \revisionK{distortion} of the axial current density as a function of chemical potential.
%        At \(\sqrt{\mu^2-M^2}\) of the order of \(R^{-1}\) the oscillation amplitude \(A\) may be fitted by the following dependence
%        \begin{equation}
%            A \approx \frac{\Omega\sqrt{\mu^2-M^2}T}{\sinh(\pi^2RT)}
%        \end{equation}
%        and its ``period'' over \(\mu\) may be estimated as
%        \begin{equation}
%            \Delta\mu \approx \frac{2}{R}
%        \end{equation}
        On Fig. \ref{fig:osc} the axial current density at $r=0$ (given in the units of \(R^{-3}\)) for the system of massless fermions is represented as a function of the chemical potential \(\mu\) (given in the units of \(R^{-1}\)).
        Temperature is taken equal to  \(T=0.1R^{-1}\) and \(\Omega=0.2R^{-1}\).
    \revisionK{
    The narrow peaks at \(\mu\approx q_{0l}+\frac{\Omega}{2},\,l\in \mathbb{N}\) explicitly reveal the spectrum.
    }
        }

\end{enumerate}

\section{Conclusions}

In the present paper the chiral vortical effect was considered taking into account the finite mass of the fermions and the finite size of the system. In addition in the infinite volume limit at zero temperature we briefly discuss possible influence of interactions.
We assume the axial symmetry and adopt the MIT boundary conditions.
The spectrum of the system was calculated and the corresponding solutions of the free Dirac equation that satisfy the given boundary conditions were found following \cite{chernodub}. The axial current density was calculated numerically.

The presence of mass modifies this density. We expect, that for the interacting massive particles the modification should be even more significant as predicted in \cite{Corr1,Corr2}. It appears that the finite size also affects essentially the axial current. In particular, at low enough temperatures the oscillations appear in the dependence of the axial current density on the chemical potential.   At any values of temperature the axial current density varies fast when the distance to the rotation axis is changed. The previous results obtained in the infinite volume limit are reproduced in our calculations only in the small vicinity of the rotation axis. 

%The increase of mass leads to the suppression of the  axial current in the bulk, but the value of the current on the boundary at the same time is increased.

Comparison of the two distinct approaches to the definition of rotation demonstrates, that at small enough values of temperature $T$ (much smaller than $|\mu - M|$) those two approaches give the same value of the axial current density in the infinite volume limit. At the same time the definition of rotation of Sect. \ref{SecFree} clearly differs from that of Sect. \ref{SectOM} for the vanishing chemical potential (i.e. at $\mu \ll T$). Namely, according to Sect. \ref{SecFree} in this case the axial current is equal to zero while the definition of Sect. \ref{SectOM} gives the non - vanishing values of the axial current. This discrepancy demonstrates, that the notion of macroscopic motion may be introduced to the quantum field theory in a variety of different ways.

The results obtained using the definition of Sect. \ref{SecFree} ensure that the coefficient at the term $\sim \mu^2$ in Eq. (\ref{eq1}) is topologically protected at vanishing temperature. That means that it is not changed when we modify smoothly the system. The introduction of interactions being such a smooth change cannot renormalize, therefore, this coefficient. There is no such a correspondence between the two mentioned approaches at large enough temperature, which is in accordance with the conclusions of \cite{Corr1,Corr2} that the corresponding term in Eq. (\ref{eq1}) may be modified due to interactions. 

The expected experimental observation of chiral vortical effect is related to the physics of heavy ion collisions. The anatomy of the chiral vortical effect discussed here is relevant for its description. In particular, the fireballs that appear during the collisions have the finite size. Therefore, the finite volume modifications of Eq. (\ref{eq1}) discussed in the present paper should  be taken into account in the consideration of the chiral vortical effect inside the rotated fireballs.
   
M.A.Z. is grateful for the useful comments to O.V.Teryaev. M.A.Z. and Z.V.K. kindly acknowledge useful discussions with M.N.Chernodub. R.A.A. is grateful for the discussions to V.I.Zakharov. The work of Z.V.K. was supported by Russian Science Foundation Grant No 16-12-10059.

\end{document}